\magnification=\magstep0
\font\bbf=cmbx10 scaled\magstep1
\hsize=6.45truein

\topglue 2in
\centerline{\bbf THREE-BODY MATRIX ELEMENTS FOR CALCULATIONS OF}
\centerline{\bbf MEAN FIELD AND exp(S) GROUND STATE CORRELATIONS.%
  \footnote*{{\it work supported by DOE grant DE-FG02-87ER-40371}}}
\centerline{by}
\centerline{Bogdan~Mihaila and Jochen~H.~Heisenberg}
\centerline{\it Dept. of Physics, University of New Hampshire,
Durham, N.H. 03824}
\vskip 0.5in

\noindent {\it Abstract.} In this document we present our approach to
the computation of three-body matrix elements, based on the Urbana
family of three-nucleon potentials.  The calculations refer only to the
necessary matrix elements needed to include the three-nucleon
interaction in the manner presented in reference~[1].

\vskip 0.5in

UNH-NPG-99-vtni; nucl-th/9912024

\vfill
\eject

\beginsection{1. Three-Nucleon Interaction V$^{tni}$.}

In the description of the Urbana potential series~[2], 
the three-nucleon interaction is presented as a sum of a long-range 
part derivable from the two-pion exchange diagrams, and a phenomenological 
short-range part, adjusted to reproduce the binding energy of the three-body 
nuclear system. 
The two-pion exchange interaction ($V_{2\pi, tni}$) is given as
$$\eqalign{ V_{2\pi, tni} = 
\sum_{cycl.}~&A_{2\pi} \lbrace \tau_1\cdot,\tau_2,\tau_1\cdot \tau_3
\rbrace \lbrace S_{12}T(r_{12})+\sigma_1\cdot\sigma_2Y(r_{12})),
                S_{13}T(r_{13})+\sigma_1\cdot\sigma_3Y(r_{13}))\rbrace\cr
&C_{2\pi}[\tau_1\cdot,\tau_2,\tau_1\cdot \tau_3]
        [S_{12}T(r_{12})+\sigma_1\cdot\sigma_2Y(r_{12})),
         S_{13}T(r_{13})+\sigma_1\cdot\sigma_3Y(r_{13}))]}\eqno(1.1)$$
Here $\sum_{cycl.}$ represents a sum over cyclic permutations over the indices
1,2, and 3. $\tau$, $\sigma$, and $S_{ij}$ are the isospin, spin, and tensor
operators, and $\lbrace,\rbrace$ and $[,]$ denote the anticommutators and
commutators. The $T(r)$ and $Y(r)$ are radial functions associated with the
tensor and Yukawa parts of the one-pion-exchange interaction, and $C_{2\pi}=
{1\over{4}}A_{2\pi}$.

The short range repulsion is phenomenological and is given as
$$V_{R,tni}=U_0\sum_{cycl.}T^2(r_{12})T^2(r_{13})\eqno(1.2)$$
{\it Note.} For the Urbana IX potential,
the fitted parameters are: $A_{2\pi}=-0.0293$ and $U_0=+0.0048$.

The relevant matrix elements for our calculation are of two types only. 
The matrix elements of the
form $V^{tni,a}_{h,a_1,a_2;h,b_1,b_2}$ are derived in section 2, whereas
the matrix elements of the form $-V^{tni,a}_{h,a_1,a_2;b_1,h,b_2}$ are
presented in section 3.

\beginsection{2. Density-Dependent Matrix Elements.}

In this section we work out the computation of the integrals
$$V^{tni}_{h,a_1,b_1,p,a_2,b_2}=
 <\phi_{h}(1)\phi_{a_1}(2)\phi_{b_1}(3)|V^{tni}|
   \phi_{p}(1)\phi_{a_2}(2)\phi_{b_2}(3)>.\eqno{(2.1)}$$
We employ the same methods as developed for the calculations of the
two-body matrix elements, namely using
Fourier-transforms in order to separate the variables. 
We use an expansion into Harmonic oscillator
functions $H^{\ell}_n(r)$ for which the Fourier transforms are
again Harmonic oscillator functions $H^{\ell}_n(q)$.

We will make use of the particular angular momentum coupling in order to
make the computations feasible.
For the matrix elements given by (2.1) and requiring $j_h=j_p$ the
angular momentum coupling is similar to that of our two-body
matrix elements:
$$\eqalign{
<(a_1\bar a_2)_{\lambda}|V^{eff}|(b_2\bar b_1)_{\lambda}>=
&\delta_{m_{h},m_{p}}
~(-)^{k_{a_1}+k_{b_2}}(-)^{(k_{a_2}+k_{b_1}+m_{a_2}-m_{b_1})}\cr
& <j_{a_1}m_{a_1}j_{a_2}-m_{a_2}|\lambda\mu>
<j_{b_2}m_{b_2}j_{b_1}-m_{b_1}|\lambda\mu>
V_{m_hm_{a_1}m_{b_1},m_pm_{a_2}m_{b_2}}\cr
<(a_1\bar a_2)_{\lambda}|V^{eff,x}|(b_2\bar b_1)_{\lambda}>=&
\delta_{m_{h},m_{p}}
~(-)^{k_{a_1}+k_{b_2}}(-)^{(k_{a_2}+k_{b_1}+m_{a_2}-m_{b_1})}\cr
& <j_{a_1}m_{a_1}j_{a_2}-m_{a_2}|\lambda\mu>
<j_{b_2}m_{b_2}j_{b_1}-m_{b_1}|\lambda\mu>
(-)V_{m_hm_{a_1}m_{b_1},m_{a_2}m_pm_{b_2}}\cr
}\eqno(2.2)$$
which includes the ``Ring''-phase for two-body $ph$-matrix elements.
Here we do sum over $m_h=m_p$ and we do sum over all other m's.
This angular momentum coupling applies as well for the matrix
elements of section 3. In this section we handle the sum over the
cyclic permutations by calculating the three matrix elements
separately
$$\eqalign{
V^{tni}_{h,a_1,b_1,p,a_2,b_2}=&
 <\phi_{h}(1)\phi_{a_1}(2)\phi_{b_1}(3)|V^{tni}|
   \phi_{p}(1)\phi_{a_2}(2)\phi_{b_2}(3)>\cr
& <\phi_{h}(2)\phi_{a_1}(3)\phi_{b_1}(1)|V^{tni}|
   \phi_{p}(2)\phi_{a_2}(3)\phi_{b_2}(1)>\cr
& <\phi_{h}(3)\phi_{a_1}(1)\phi_{b_1}(2)|V^{tni}|
   \phi_{p}(3)\phi_{a_2}(1)\phi_{b_2}(2)>\cr}
\eqno{(2.3)}$$

\beginsection{2a. Short range repulsion term.}

As this term does not contain any additional operators it is
particularly simple to handle.
We use Eq. (2.2) of Reference~[3]
$$T^2(r_{12})=4\pi~{2\over{\pi}} \int q^2dq \tilde T(q)
\sum_{\ell} (-)^{\ell} \hat \ell ~j_{\ell}(qr_1)
j_{\ell}(qr_2) \bigl[Y_{\ell}(\hat r_1)\otimes Y_{\ell}(\hat
r_2)\bigr]^{(0)}\eqno{(2.4)}$$
using
$$\tilde T(q)=\int r_{12}^2dr_{12}T^2(r_{12})j_0(qr_{12})\eqno{(2.5)}$$
Correspondingly we write
$$\eqalign{
T^2(r_{12})T^2(r_{13})=&(4\pi)^2\sum_{\ell_2,\ell_3}
(-)^{(\ell_2+\ell_3)}\hat \ell_2 \hat \ell_3 \cr
\times & {2\over{\pi}}\int q_2^2dq_2 \tilde T(q_2) 
j_{\ell_2}(q_2r_1) j_{\ell_2}(q_2r_2)
\times  {2\over{\pi}}\int q_3^2dq_3 \tilde T(q_3) 
j_{\ell_3}(q_3r_1) j_{\ell_3}(q_3r_3)\cr
\times & \bigl[Y_{\ell_2}(\hat r_1)\otimes Y_{\ell_2}(\hat
r_2)\bigr]^{(0)}\bigl[Y_{\ell_3}(\hat r_1)\otimes Y_{\ell_3}(\hat
r_3)\bigr]^{(0)} }\eqno{(2.6)}$$
recoupling the spherical harmonics results in
$$\eqalign{
T^2(r_{12})T^2(r_{13})=&(4\pi)^2{1\over{\sqrt{4\pi}}}
\sum_{\ell_1,\ell_2,\ell3}
(-)^{(\ell_2+\ell_3)}\hat \ell_2 \hat \ell_3 \cr
\times & {2\over{\pi}}\int q_2^2dq_2 \tilde T(q_2) 
j_{\ell_2}(q_2r_1) j_{\ell_2}(q_2r_2)
\times  {2\over{\pi}}\int q_3^2dq_3 \tilde T(q_3) 
j_{\ell_3}(q_3r_1) j_{\ell_3}(q_3r_3)\cr
\times & \Bigl[Y_{\ell_1}(\hat r_1)\otimes \bigl[Y_{\ell_2}(\hat
r_2)\otimes Y_{\ell_3}(\hat r_3)\bigr]^{(\ell_1)}\Bigr]^{(0)} 
}\eqno{(2.7)}$$
This leads to $\ell_1=0$ for the first matrix element, $\ell_2=0$ for
the second and $\ell_3=0$ for the third.
In turn we can write the interaction for the three matrix elements
respectively as
$$\eqalign{
T^2(r_{12})T^2(r_{13})=&
\sum_{\ell}
(-)^{\ell}\hat \ell  
~\bigl[ \sqrt{4\pi}Y_{\ell}(\hat r_2)\odot  
\sqrt{4\pi}Y_{\ell}(\hat r_3)\bigr]\cr
\times & {2\over{\pi}}
  \int q_2^2dq_2 \tilde T(q_2) j_{\ell}(q_2r_1) j_{\ell}(q_2r_2)
\times {2\over{\pi}}
  \int q_3^2dq_3 \tilde T(q_3) j_{\ell}(q_3r_1) j_{\ell}(q_3r_3)
 }\eqno{(2.8)}$$
for the first matrix element and
$$\eqalign{
T^2(r_{12})T^2(r_{13})=&
\sum_{\ell}
~\bigl[ \sqrt{4\pi}Y_{\ell}(\hat r_1)\odot  
\sqrt{4\pi}Y_{\ell}(\hat r_3)\bigr]\cr
\times & {2\over{\pi}}
  \int q_2^2dq_2 \tilde T(q_2) j_{0}(q_2r_1) j_{0}(q_2r_2)
\times {2\over{\pi}}
  \int q_3^2dq_3 \tilde T(q_3) j_{\ell}(q_3r_1) j_{\ell}(q_3r_3)\cr
T^2(r_{12})T^2(r_{13})=&
\sum_{\ell}
~\bigl[ \sqrt{4\pi}Y_{\ell}(\hat r_1)\odot  
\sqrt{4\pi}Y_{\ell}(\hat r_2)\bigr]\cr
\times & {2\over{\pi}}
  \int q_3^2dq_3 \tilde T(q_3) j_{0}(q_3r_1) j_{0}(q_3r_3)
\times {2\over{\pi}}
  \int q_2^2dq_2 \tilde T(q_2) j_{\ell}(q_2r_1) j_{\ell}(q_2r_2)
 }\eqno{(2.9)}$$
for the second and third respectively.

For the first matrix element we write
$$\eqalign{
<\phi_{h}(1)\phi_{a_1}(2)\phi_{b_1}(3)&|T^2(r_{12})T^2(r_{13})|
   \phi_{p}(1)\phi_{a_2}(2)\phi_{b_2}(3)>=
  (-)^{\lambda}\hat\lambda
   \int r_1^2dr_1R_{h}(r_1)R_{p}(r_1)\cr
\times & (-)^{k_{a_1}}{{\sqrt{4\pi}}\over{\hat \lambda}}
<j_{a_1}\Vert Y_{\lambda}
  \Vert j_{a_2}>\int r_2^2dr_2R_{a_1}(r_2)R_{a_2}(r_2){2\over{\pi}}
 \int q_2^2dq_2 \tilde T(q_2) j_{\lambda}(q_2r_1) j_{\lambda}(q_2r_2)\cr
\times & (-)^{k_{b_2}}{{\sqrt{4\pi}}\over{\hat \lambda}}
<j_{b_2}\Vert Y_{\lambda}
  \Vert j_{b_1}>\int r_3^2dr_3R_{b_1}(r_3)R_{b_2}(r_3){2\over{\pi}}
 \int q_3^2dq_3 \tilde T(q_3) j_{\lambda}(q_3r_1) j_{\lambda}(q_3r_3)
}\eqno{(2.10)}$$
We now assume $R_{a_1}(r)R_{a_2}(r)$ is expanded into harmonic
oscillator functions as
$$R_{a_1}(r_2)R_{a_2}(r_2)=\sum_n A^a_n H^{\lambda}_n(r_2)\eqno{(2.11)}$$
Using Gauss integration the coefficients can be written in terms of the
functions at the Gauss-points, $r_i$, with weights, $w_i$, as
$$A^a_n=\sum_i \bar R_{a_1}(r_i)\bar R_{a_2}(r_i)
   \bar H^{\lambda}_n(r_i)r_i^2w_i\eqno{(2.12)}$$
where the barred functions are those without the exponentials. In turn,
we can write the integral
$$\eqalign{
\int r_2^2dr_2R_{a_1}(r_2)R_{a_2}(r_2){2\over{\pi}}
& \int q_2^2dq_2 \tilde T(q_2) j_{\lambda}(q_2r_1) j_{\lambda}(q_2r_2)=
\cr & \sum_{n,m} A^a_n T^{\lambda}_{m,n}H^{\lambda}_m(r_1)}\eqno{(2.13)}$$
where 
$$T^{\lambda}_{m,n}=\int q_2^2dq_2 H^{\lambda}_m(q_2) \tilde T(q_2)
                 H^{\lambda}_n(q_2)
=\sum_s q_s^2w_s \bar H^{\lambda}_m(q_s) \tilde T(q_s)
 \bar H^{\lambda}_n(q_s)\eqno{(2.14)}$$
We proceed similarly with the integration over $q_3$ and $r_3$.
Thus this matrix element can be written as
$$\eqalign{
<\phi_{h}(1)\phi_{a_1}(2)\phi_{b_1}(3)&|T^2(r_{12})T^2(r_{13})|
   \phi_{p}(1)\phi_{a_2}(2)\phi_{b_2}(3)>=\cr
(-)^{k_{b_1}}&{{\sqrt{4\pi}}\over{\hat \lambda}}<j_{a_1}\Vert Y_{\lambda}
  \Vert j_{a_2}>
(-)^{k_{b_2}}{{\sqrt{4\pi}}\over{\hat \lambda}}<j_{b_2}\Vert Y_{\lambda}
  \Vert j_{b_1}>\cr
\times & \sum_{i,j}\bar R_{a_1}(r_i)\bar R_{a_2}(r_i)
  \bar R_{b_1}(r_j)\bar R_{b_2}(r_j) \Delta S_{i,j}
}\eqno{(2.15)}$$
where
$$\Delta S_{i,j}=\bar H^{\lambda}_n(r_i)r_i^2w_i
                 \bar H^{\lambda}_k(r_j)r_j^2w_j
                 T^{\lambda}_{n,m}T^{\lambda}_{k,l}D^{\lambda}_{m,l}
   \eqno{(2.16)}$$
with
$$D^{\lambda}_{m,l}=(-)^{\lambda}\hat \lambda
\int  r_s^2 dr_s H^{\lambda}_m(r_s)  H^{\lambda}_l(r_s) 
R_{p}(r_s)R_{h}(r_s)
=(-)^{\lambda}\hat \lambda\sum_sr_s^2 w_s \bar H^{\lambda}_m(r_s)
\bar H^{\lambda}_l(r_s)R_{p}(r_s)R_{h}(r_s)
\eqno{(2.17)}$$

For the second matrix element we obtain
$$\eqalign{
<\phi_{b_1}(1)\phi_{h}(2)\phi_{a_1}(3)&|T^2(r_{12})T^2(r_{13})|
   \phi_{b_2}(1)\phi_{p}(2)\phi_{a_2}(3)>=
   \int r_2^2dr_2R_{h}(r_2)R_{p}(r_2)\cr
\times & (-)^{k_{a_1}}{{\sqrt{4\pi}}\over{\hat \lambda}}
<j_{b_1}\Vert Y_{\lambda}
  \Vert j_{b_2}>\int r_1^2dr_1R_{b_1}(r_1)R_{b_2}(r_1){2\over{\pi}}
 \int q_2^2dq_2 \tilde T(q_2) j_{0}(q_2r_1) j_{0}(q_2r_2)\cr
\times & (-)^{k_{a_2}}{{\sqrt{4\pi}}\over{\hat \lambda}}
<j_{a_2}\Vert Y_{\lambda}
  \Vert j_{a_1}>\int r_3^2dr_3R_{a_1}(r_3)R_{a_2}(r_3){2\over{\pi}}
 \int q_3^2dq_3 \tilde T(q_3) j_{\lambda}(q_3r_1) j_{\lambda}(q_3r_3)}
\eqno{(2.18)}$$
Again, this equivalent two-body matrix element takes the form
$$\eqalign{
<\phi_{b_1}(1)\phi_{h}(2)\phi_{a_1}(3)&|T^2(r_{12})T^2(r_{13})|
   \phi_{b_2}(1)\phi_{p}(2)\phi_{a_2}(3)>=\cr
(-)^{k_{b_1}}&{{\sqrt{4\pi}}\over{\hat \lambda}}<j_{b_1}\Vert Y_{\lambda}
  \Vert j_{b_2}>
(-)^{k_{a_2}} {{\sqrt{4\pi}}\over{\hat \lambda}}<j_{a_2}\Vert Y_{\lambda}
  \Vert j_{a_1}>\cr
\times & \sum_{i,j}\bar R_{b_1}(r_i)\bar R_{b_2}(r_i)
  \bar R_{a_1}(r_j)\bar R_{a_2}(r_j) \Delta S_{i,j}
}\eqno{(2.19)}$$
In this case
$$\Delta S_{i,j}=r_i^2w_ir_j^2w_jH^0_m(r_i)T^0_{m,n} C^0_n
\bar H^{\lambda}_l(r_j)\bar H^{\lambda}_k(r_i)T^{\lambda}_{k,l}
\eqno{(2.20)}$$
where 
$$C^0_n=\int r_2^2dr_2 R_{h}(r_2)R_{p}(r_2)H^0_n(r_2)
=\sum_k r_k^2w_k \bar R_{h}(r_k)\bar R_{p}(r_k)\bar H^0_n(r_k)
\eqno{(2.21)}$$

Similarly, the third density term is
$$\eqalign{
T^2(r_{12})T^2(r_{13})=&\Bigl({1\over{\pi}}\Bigr)^3
\sum_{\ell}
~\bigl[ Y_{\ell}(\hat r_1)\odot  Y_{\ell}(\hat r_2)\bigr]\cr
\times & \int q_2^2dq_2 \tilde T(q_2) j_{\ell}(q_2r_1) j_{\ell}(q_2r_2)
\times  \int q_3^2dq_3 \tilde T(q_3) j_{0}(q_3r_1) j_{0}(q_3r_3)
 }\eqno{(2.22)}$$
The form can be obtained from the previous case by using the
transpose i.e. $\Delta S^c_{i,j}=\Delta S^b_{j,i}$.

\beginsection{2b. Two-pion exchange term.}

For the model V $2\pi$-exchange TNI takes the form
$$\eqalign{
V^{tni,a}_{2\pi}=&A_{2\pi}\lbrace\tau_1\tau_2,\tau_1\tau_3\rbrace
 \lbrace (S_{12}T(r_{12})+\sigma_1\sigma_2 Y(r_{12})),
         (S_{13}T(r_{13})+\sigma_1\sigma_3 Y(r_{13}))\rbrace\cr
}\eqno{(2.23)}$$
We can write the tensor interaction as
$$S_{12}=3 <1010|K0>\sqrt{4\pi}\Bigl[Y^{(K)}(\hat r_{12})
  \otimes \bigl[ \sigma_1 \otimes \sigma_2 \bigr]^{(K)}\Bigr]^{(0)}
\eqno{(2.24)}$$
with $K=2$, whereas the sigma interaction is identical with $K=0$.
By defining the form factors
$$ 4\pi~V^K(q)=4\pi\int V^K(r)j_K(qr)r^2dr\eqno{(2.25)}$$
with $V(r)=Y$ for $K=0$ and $V(r)=T$ for $K=2$
the interaction can be written as 
$$\eqalign{A_{2\pi}&(4\pi)^2
\sum_{K_2}\sum_{K_3}<1010|K_20><1010|K_30>{{9}\over{\hat K_2\hat K_3}}
<\ell_10\ell_20|K_20><\ell_30\ell_40|K_30>\cr
&{2\over{\pi}}\int q_2^2dq_2 V^{K_2}(q_2)\hat \ell_1\hat\ell_2
(i)^{(\ell_1-\ell_2-K_2)}j_{\ell_1}(q_2r_1)j_{\ell_2}(q_2r_2)\cr
&{2\over{\pi}}\int q_3^2dq_3 V^{K_3}(q_3)\hat \ell_4\hat\ell_3
(i)^{(\ell_4-\ell_3-K_3)}j_{\ell_4}(q_3r_1)j_{\ell_3}(q_3r_3)\cr
&\biggl[\bigl[Y_{\ell_1}(\hat r_1)\otimes Y_{\ell_2}(\hat r_2)\bigr]
  ^{(K_2)}\otimes
  \bigl[\sigma_1\otimes \sigma_2\bigr]^{(K_2)}\biggr]^{(0)}
  \biggl[\bigl[Y_{\ell_4}(\hat r_1)\otimes Y_{\ell_3}(\hat r_3)\bigr]
  ^{(K_3)}\otimes
  \bigl[\sigma_1\otimes \sigma_3\bigr]^{(K_3)}\biggr]^{(0)}
}\eqno{(2.26)}$$
Here $K_2$ or $K_3$ are either 0, for the $\sigma\cdot\sigma$-term
or 2, for the tensor term.  These need to be recoupled:
$$\eqalign{
\biggl[\bigl[Y_{\ell_1}(\hat r_1)&\otimes Y_{\ell_2}(\hat r_2)\bigr]
  ^{(K_2)}\otimes
  \bigl[\sigma_1\otimes \sigma_2\bigr]^{(K_2)}\biggr]^{(0)}
  \biggl[\bigl[Y_{\ell_4}(\hat r_1)\otimes Y_{\ell_3}(\hat r_3)\bigr]
  ^{(K_3)}\otimes
  \bigl[\sigma_1\otimes \sigma_3\bigr]^{(K_3)}\biggr]^{(0)}\cr
=(-)^{\ell_2+\lambda_2}&(-)^{\ell_3+\lambda_3}\hat K_2\hat \lambda_2
\hat K_3\hat \lambda_3 
\Bigl\lbrace\matrix{\ell_1&\ell_2&K_2\cr 1&1&\lambda_2}\Bigr\rbrace
\Bigl\lbrace\matrix{\ell_4&\ell_3&K_3\cr 1&1&\lambda_3}\Bigr\rbrace\cr
&\biggl[\bigl[Y_{\ell_1}(\hat r_1)\otimes\sigma_1\bigr]^{(\lambda_2)}
      \otimes
       \bigl[Y_{\ell_2}(\hat r_2)\otimes\sigma_2\bigr]^{(\lambda_2)}\biggr]
     ^{(0)}\biggl[\bigl[Y_{\ell_4}(\hat
     r_1)\otimes\sigma_1\bigr]^{(\lambda_3)}\otimes \bigl[Y_{\ell_3}(\hat
     r_3)\otimes\sigma_3\bigr]^{(\lambda_3)}\biggr]^{(0)}
}\eqno{(2.27)}$$
This, in turn we can write as
$$\eqalign{
=(-)^{\ell_2+\lambda_2}(-)^{\ell_3+\lambda_3}\hat K_2\hat \lambda_2
\hat K_3\hat \lambda_3 &{\hat J\over{\hat \lambda_2\hat\lambda_3}}
\Bigl\lbrace\matrix{\ell_1&\ell_2&K_2\cr 1&1&\lambda_2}\Bigr\rbrace
\Bigl\lbrace\matrix{\ell_4&\ell_3&K_3\cr 1&1&\lambda_3}\Bigr\rbrace\cr
\Biggl[\biggl[\bigl[Y_{\ell_1}(\hat r_1)\otimes\sigma_1\bigr]^{(\lambda_2)}
    & \otimes  \bigl[Y_{\ell_4}(\hat r_1)\otimes\sigma_1\bigr]
   ^{(\lambda_3)}\biggr]^{(J)} \otimes
     \biggl[\bigl[Y_{\ell_2}(\hat
     r_2)\otimes\sigma_2\bigr]^{(\lambda_2)} \otimes \bigl[Y_{\ell_3}(\hat
     r_3)\otimes\sigma_3\bigr]^{(\lambda_3)}\biggr]^{(J)}\Biggr]^{(0)}\cr
=(-)^{\ell_2+\lambda_2+\ell_3+\lambda_3} \hat K_2 \hat K_3 \hat
\lambda_2 \hat \lambda_3 \hat J & \hat L\hat S\Biggl\lbrace
\matrix{\ell_1 & 1 & \lambda_2\cr \ell_4& 1& \lambda_3\cr
L & S & J}\Biggr\rbrace
\Bigl\lbrace\matrix{\ell_1&\ell_2&K_2\cr 1&1&\lambda_2}\Bigr\rbrace
\Bigl\lbrace\matrix{\ell_4&\ell_3&K_3\cr 1&1&\lambda_3}\Bigr\rbrace\cr
\Biggl[\biggl[\bigl[Y_{\ell_1}(\hat r_1)\otimes Y_{\ell_4}(\hat r_1)\bigr]^{L}
    & \otimes  \bigl[\sigma_1\otimes\sigma_1\bigr]
   ^{(S)}\biggr]^{(J)} \otimes
     \biggl[\bigl[Y_{\ell_2}(\hat
     r_2)\otimes\sigma_2\bigr]^{(\lambda_2)}\otimes \bigl[Y_{\ell_3}(\hat
     r_3)\otimes\sigma_3\bigr]^{(\lambda_3)}\biggr]^{(J)}\Biggr]^{(0)}
}\eqno{(2.28)}$$
$$\eqalign{=
(-)^{\ell_2+\lambda_2+\ell_3+\lambda_3} \hat K_2 \hat K_3 \hat
\lambda_2 \hat \lambda_3 \hat J & \hat \ell_1 \hat \ell_4 \hat S
{1\over{\sqrt{4\pi}}} \Biggl\lbrace
\matrix{\ell_1 & 1 & \lambda_2\cr \ell_4& 1& \lambda_3\cr
L & S & J}\Biggr\rbrace
\Bigl\lbrace\matrix{\ell_1&\ell_2&K_2\cr 1&1&\lambda_2}\Bigr\rbrace
\Bigl\lbrace\matrix{\ell_4&\ell_3&K_3\cr 1&1&\lambda_3}\Bigr\rbrace\cr
\Biggl[\biggl[Y_{L}(\hat r_1)
    & \otimes  \bigl[\sigma_1\otimes\sigma_1\bigr]
   ^{(S)}\biggr]^{(J)} \otimes
     \biggl[\bigl[Y_{\ell_2}(\hat
     r_2)\otimes\sigma_2\bigr]^{(\lambda_2)}\otimes \bigl[Y_{\ell_3}(\hat
     r_3)\otimes\sigma_3\bigr]^{(\lambda_3)}\biggr]^{(J)}\Biggr]^{(0)}
}\eqno{(2.29)}$$
As the reduced one-body matrix element of $(Y_{\ell}\otimes \sigma)^{\lambda}$
vanishes for $\lambda=0$, only the first matrix element in (2.2) gives a
contribution. For it we have
 $S=0$ and $L=0$. This implies $J=0$ and
$\lambda_2=\lambda_3$ as well as $\ell_1=\ell_4$. 
This term leads to a density dependent finite range Migdal g
or tensor
interaction. For this case we evaluate the 9-j symbol and simplify
the interaction to
$$
{1\over{4\pi}}(-)^{(\ell_2+\ell_3+\lambda+1)}
\hat K_2 \hat K_3 \hat \ell_1 
\Bigl\lbrace\matrix{\ell_1&\ell_2&K_2\cr 1&1&\lambda}\Bigr\rbrace
\Bigl\lbrace\matrix{\ell_1&\ell_3&K_3\cr 1&1&\lambda}\Bigr\rbrace
\Bigl[\bigl[Y_{\ell_2}(\hat r_2)\otimes\sigma_2\bigr]^{(\lambda)}\odot
\bigl[Y_{\ell_3}(\hat r_3)\otimes\sigma_3\bigr]^{(\lambda)}\Bigr]^{(0)}
\eqno{(2.30)}$$
By using Eq. (2.1) of Reference~[3] we write the matrix element as
$$\eqalign{
<(a_1\bar a_2)_{\lambda}&|V^{eff}|(b_2\bar b_1)_{\lambda}>=
9A_{2\pi}\cr
&\sum_{K_2=0,2}\sum_{K_3=0,2} \hat \ell_1\hat \ell_2
\hat \ell_1\hat \ell_3 <1010|K_20><1010|K_30>
<\ell_1 0\ell_2 0|K_20><\ell_1 0\ell_3 0|K_30>\cr
&(i)^{(\ell_2-K_2+\ell_3-K_3)}(-)^{(\ell_1+\lambda+1)}\hat \ell_1
\Bigl\lbrace\matrix{\ell_1&\ell_2&K_2\cr 1&1&\lambda}\Bigr\rbrace
\Bigl\lbrace\matrix{\ell_1&\ell_3&K_3\cr 1&1&\lambda}\Bigr\rbrace\cr
&\int r_1^2dr_1 R_{h}(r_1)R_{p}(r_1)
{{\sqrt{4\pi}}\over{\hat \lambda}}(-)^{k_{a_1}}
<j_{a_1}\Vert [Y^{(\ell_2)}\sigma]^{\lambda} \Vert j_{a_2}>
{{\sqrt{4\pi}}\over{\hat \lambda}}(-)^{k_{b_2}}
<j_{b_2}\Vert [Y^{(\ell_3)}\sigma]^{\lambda} \Vert j_{b_1}>
\cr
&{2\over{\pi}}\int q_2^2dq_2 V^{K_2}(q_2)\int r_2^2dr_2
j_{\ell_1}(q_2r_1)j_{\ell_2}(q_2r_2) R_{a_1}(r_2)R_{a_2}(r_2)\cr
&{2\over{\pi}}\int q_3^2dq_3 V^{K_3}(q_3)\int r_3^2dr_3
j_{\ell_1}(q_3r_1)j_{\ell_3}(q_3r_3) R_{b_1}(r_3)R_{b_2}(r_3)}
\eqno{(2.31)}$$
Exchanging variables 2 and 3 results in the identical expression.
Thus, the commutator vanishes for this density dependent term, whereas
the anti-commutator obtains a factor of 2. Further, we use
$\lbrace \vec\tau_1\vec\tau_2,\vec\tau_1\vec\tau_3\rbrace=
2\vec\tau_2\vec\tau_3$.
Again, we write the integrals as
sums over Gauss-points
$$\eqalign{
<(a_1\bar a_2)_{\lambda}&|V^{eff}|(b_2\bar b_1)_{\lambda}>=
4\cdot 9A_{2\pi}~<\vec\tau_2\vec\tau_3>\cr
&\sum_{K_2=0,2}\sum_{K_3=0,2} \hat \ell_1\hat \ell_2
\hat \ell_1\hat \ell_3  <1010|K_20><1010|K_30>
<\ell_1 0\ell_2 0|K_20><\ell_1 0\ell_3 0|K_30>\cr
&(i)^{(\ell_2-K_2+\ell_3-K_3)}(-)^{(\lambda+1)}
\Bigl\lbrace\matrix{\ell_1&\ell_2&K_2\cr 1&1&\lambda}\Bigr\rbrace
\Bigl\lbrace\matrix{\ell_1&\ell_3&K_3\cr 1&1&\lambda}\Bigr\rbrace\cr
&
{{\sqrt{4\pi}}\over{\hat \lambda}}(-)^{k_{a_1}}
<j_{a_1}\Vert [Y^{(\ell_2)}\sigma]^{\lambda} \Vert j_{a_2}>
{{\sqrt{4\pi}}\over{\hat \lambda}}(-)^{k_{b_2}}
<j_{b_2}\Vert [Y^{(\ell_3)}\sigma]^{\lambda} \Vert j_{b_1}>
\cr
&D^{\ell_1}_{m,s}V^{K_2,\ell_1\ell_2}_{m,n}
V^{K_3,\ell_1\ell_3}_{s,t}
\bar H^{\ell_2}_n(r_i)r_i^2w_i\bar H^{\ell_3}_t(r_j)r_j^2w_j
~~~\bar R_{a_1}(r_i)\bar R_{a_2}(r_i)\bar R_{b_1}(r_j)
   \bar R_{b_2}(r_j)
}\eqno{(2.32)}$$
where we have used the same definition of $D$ as before (3.17)
and
$$V^{K,\ell_1\ell_2}_{m,n}=\sum q_k^2w_k \bar H^{\ell_1}_m(q_k)
  \bar H^{\ell_2}_n(q_k)V^{K}(q_k)\eqno{(2.33)}$$
The selection rules require $\ell_1+\ell_2=even$ and
$\ell_1+\ell_3=even$ which in turn requires $\ell_2+\ell_3=even$.
Further, $\ell_1,\ell_2,\ell_3$ are all restricted to values
of $\lambda,\lambda\pm 1$.

We now discuss the four cases separately. The first case, $K_2=0$,$K_3=0$
leads to a density dependent $\sigma_1\sigma_2$ interaction:
$$\eqalign{
<(a_1\bar a_2)_{\lambda}&|V^{eff,\sigma}|(b_2\bar b_1)_{\lambda}>=
4A_{2\pi}~<\vec\tau_2\vec\tau_3>
D^{\ell}_{m,s}V^{0,\ell\ell}_{m,n}
V^{0,\ell\ell}_{s,t}
\bar H^{\ell}_n(r_i)r_i^2w_i\bar H^{\ell}_t(r_j)r_j^2w_j\cr
& (-)^{(\ell+\lambda+1)}
{{\sqrt{4\pi}}\over{\hat \lambda}}(-)^{k_{a_1}}
<j_{a_1}\Vert [Y^{(\ell)}\sigma]^{\lambda} \Vert j_{a_2}>
{{\sqrt{4\pi}}\over{\hat \lambda}}(-)^{k_{b_2}}
<j_{b_2}\Vert [Y^{(\ell)}\sigma]^{\lambda} \Vert j_{b_1}>
\cr
&\bar R_{a_1}(r_i)\bar R_{a_2}(r_i)\bar R_{b_1}(r_j)
   \bar R_{b_2}(r_j)
}\eqno{(2.34)}$$
second case ($K_2=0$,$K_3=2$):
$$\eqalign{
<(a_1\bar a_2)_{\lambda}&|V^{eff,tensor}|(b_2\bar b_1)_{\lambda}>=
4 A_{2\pi}~<\vec\tau_2\vec\tau_3>~~
 \sqrt{6}\hat \ell_1\hat \ell_3 
(i)^{(\ell_3+\ell_1)}<\ell_1 0\ell_3 0|20>
\Bigl\lbrace\matrix{\ell_1&\ell_3&2\cr 1&1&\lambda}\Bigr\rbrace\cr
&D^{\ell_1}_{m,s}V^{0,\ell_1\ell_1}_{m,n}
V^{2,\ell_1\ell_3}_{s,t}
\bar H^{\ell_1}_n(r_i)r_i^2w_i\bar H^{\ell_3}_t(r_j)r_j^2w_j\cr
&
{{\sqrt{4\pi}}\over{\hat \lambda}}(-)^{k_{a_1}}
<j_{a_1}\Vert [Y^{(\ell_1)}\sigma]^{\lambda} \Vert j_{a_2}>
{{\sqrt{4\pi}}\over{\hat \lambda}}(-)^{k_{b_2}}
<j_{b_2}\Vert [Y^{(\ell_3)}\sigma]^{\lambda} \Vert j_{b_1}>
\cr
&~~\bar R_{a_1}(r_i)\bar R_{a_2}(r_i)\bar R_{b_1}(r_j)
   \bar R_{b_2}(r_j)
}\eqno{(2.35)}$$
third case ($K_2=2$,$K_3=0$):
$$\eqalign{
<(a_1\bar a_2)_{\lambda}&|V^{eff,tensor}|(b_2\bar b_1)_{\lambda}>=
4 A_{2\pi}~<\vec\tau_2\vec\tau_3>~~
\sqrt{6}\hat \ell_1\hat \ell_3
<\ell_1 0\ell_3 0|20>
(i)^{(\ell_3+\ell_1)}
\Bigl\lbrace\matrix{\ell_1&\ell_3&2\cr 1&1&\lambda}\Bigr\rbrace
\cr
&D^{\ell_3}_{m,s}V^{2,\ell_3\ell_1}_{m,n}
V^{0,\ell_3\ell_3}_{s,t}
\bar H^{\ell_1}_n(r_i)r_i^2w_i\bar H^{\ell_3}_t(r_j)r_j^2w_j\cr
&
{{\sqrt{4\pi}}\over{\hat \lambda}}(-)^{k_{a_1}}
<j_{a_1}\Vert [Y^{(\ell_1)}\sigma]^{\lambda} \Vert j_{a_2}>
{{\sqrt{4\pi}}\over{\hat \lambda}}(-)^{k_{b_2}}
<j_{b_2}\Vert [Y^{(\ell_3)}\sigma]^{\lambda} \Vert j_{b_1}>
\cr
&~~\bar R_{a_1}(r_i)\bar R_{a_2}(r_i)\bar R_{a_1}(r_j)
   \bar R_{a_2}(r_j)
}\eqno{(2.36)}$$
This term is similar to the previous term. They represent a density
dependent tensor interaction.
The last term, the tensor squared term has ($K_2=2$,$K_3=2$):
$$\eqalign{
24 A_{2\pi}&~\vec\tau_2\vec\tau_3~~
 \hat \ell_1\hat \ell_2
\hat \ell_1\hat \ell_3 
<\ell_1 0\ell_2 0|20><\ell_1 0\ell_3 0|20>\cr
&(i)^{(\ell_2+\ell_3)}(-)^{(\lambda+1)}
\Bigl\lbrace\matrix{\ell_1&\ell_2&2\cr 1&1&\lambda}\Bigr\rbrace
\Bigl\lbrace\matrix{\ell_1&\ell_3&2\cr 1&1&\lambda}\Bigr\rbrace\cr
&
{{\sqrt{4\pi}}\over{\hat \lambda}}(-)^{k_{a_1}}
<j_{a_1}\Vert [Y^{(\ell_2)}\sigma]^{\lambda} \Vert j_{a_2}>
{{\sqrt{4\pi}}\over{\hat \lambda}}(-)^{k_{b_2}}
<j_{b_1}\Vert [Y^{(\ell_3)}\sigma]^{\lambda} \Vert j_{b_2}>
\cr
&D^{\ell_1}_{m,s}V^{2,\ell_1\ell_2}_{m,n}
V^{2,\ell_1\ell_3}_{s,t}
\bar H^{\ell_2}_n(r_i)r_i^2w_i\bar H^{\ell_3}_t(r_j)r_j^2w_j
~~~\bar R_{a_1}(r_i)\bar R_{a_2}(r_i)\bar R_{b_1}(r_j)
   \bar R_{b_2}(r_j)
}\eqno{(2.37)}$$
From this case we can split up terms that are similar to the
previous ones. We write
$$\eqalign{
24A_{2\pi}&~<\vec\tau_2\vec\tau_3>~~
 \hat \ell_1\hat \ell_2
\hat \ell_1\hat \ell_3 
<\ell_1 0\ell_2 0|20><\ell_1 0\ell_3 0|20>\cr
&(i)^{(\ell_2+\ell_3)}\sum_k(-)^{(\ell_1+\ell_2+\ell_3+k)}
(2k+1)
\Bigl\lbrace\matrix{\ell_2&\ell_3&k\cr 1&1&\lambda}\Bigr\rbrace
\Bigl\lbrace\matrix{\ell_2&\ell_3&k\cr 2&2&\ell_1}\Bigr\rbrace
\Bigl\lbrace\matrix{2&2&k\cr 1&1&1}\Bigr\rbrace\cr
&
{{\sqrt{4\pi}}\over{\hat \lambda}}(-)^{k_{a_1}}
<j_{a_1}\Vert [Y^{(\ell_2)}\sigma]^{\lambda} \Vert j_{a_2}>
{{\sqrt{4\pi}}\over{\hat \lambda}}(-)^{k_{b_2}}
<j_{b_2}\Vert [Y^{(\ell_3)}\sigma]^{\lambda} \Vert j_{b_1}>
\cr
&D^{\ell_1}_{m,s}V^{2,\ell_1\ell_2}_{m,n}
V^{2,\ell_1\ell_3}_{s,t}
\bar H^{\ell_2}_n(r_i)r_i^2w_i\bar H^{\ell_3}_t(r_j)r_j^2w_j
~~~\bar R_{a_1}(r_i)\bar R_{a_2}(r_i)\bar R_{b_1}(r_j)
   \bar R_{b_2}(r_j)
}\eqno{(2.38)}$$
The term with $k=0$ can be combined with the term (2.34). Also,
the term with $k=2$ can be 
combined with the terms (2.35) and (2.36).

Thus, we end up with three terms. The first corresponds to a density
dependent $\sigma_1\sigma_2$ term and can be added directly to that term
with the form:
$$\eqalign{
<(a_1\bar a_2)_{\lambda}&|V^{eff,\sigma}|(b_2\bar b_1)_{\lambda}>=
4 A_{2\pi}~<\vec\tau_2\vec\tau_3>
\bar H^{\ell}_n(r_i)r_i^2w_i\bar H^{\ell}_t(r_j)r_j^2w_j\cr
&\Bigl[D^{\ell}_{m,s}V^{0,\ell\ell}_{m,n}
V^{0,\ell\ell}_{s,t}+\sum_{\ell_1}{2\over{5}}(2\ell_1+1)
<\ell_10\ell0|20>^2D^{\ell_1,a}_{m,s}V^{2,\ell_1\ell}_{s,t}
V^{2,\ell_1\ell}_{m,n}\Bigr]\cr
& (-)^{(\ell+\lambda+1)}
{{\sqrt{4\pi}}\over{\hat \lambda}}(-)^{k_{a_1}}
<j_{a_1}\Vert [Y^{(\ell)}\sigma]^{\lambda} \Vert j_{a_2}>
{{\sqrt{4\pi}}\over{\hat \lambda}}(-)^{k_{b_2}}
<j_{b_2}\Vert [Y^{(\ell)}\sigma]^{\lambda} \Vert j_{b_1}>
\cr
&\bar R_{a_1}(r_i)\bar R_{a_2}(r_i)\bar R_{b_1}(r_j)
   \bar R_{b_2}(r_j)
}\eqno{(2.40)}$$
The second corresponds to a density dependent tensor interaction:
$$\eqalign{
<(a_1\bar a_2)_{\lambda}&|V^{eff,tensor}|(b_2\bar b_1)_{\lambda}>=
4 A_{2\pi}~<\vec\tau_2\vec\tau_3>~~
\bar H^{\ell_1}_n(r_i)r_i^2w_i\bar H^{\ell_3}_t(r_j)r_j^2w_j\cr
&\Bigl[D^{\ell_3}_{m,s}V^{2,\ell_3\ell_1}_{m,n}
V^{0,\ell_3\ell_3}_{s,t}+
D^{\ell_1}_{m,s}V^{0,\ell_1\ell_1}_{m,n}
V^{2,\ell_1\ell_3}_{s,t}+
5\sqrt{6}\sum_{\ell}(2\ell+1)(-)^{\ell}\cr
&{{<\ell 0\ell_1 0|20><\ell 0 \ell_3 0|20>}\over{<\ell_1 0 \ell_3
0 | 20>}}
\Bigl\lbrace\matrix{\ell_1&\ell_3&2\cr 2&2&\ell}\Bigr\rbrace
\Bigl\lbrace\matrix{2&2&2\cr 1&1&1}\Bigr\rbrace
D^{\ell}_{m,s}V^{2,\ell\ell_1}_{m,n}
V^{2,\ell\ell_3}_{s,t}
\Bigr]\cr
&\sqrt{6}\hat \ell_1\hat \ell_3
<\ell_1 0\ell_3 0|20>
(i)^{(\ell_3+\ell_1)}
\Bigl\lbrace\matrix{\ell_1&\ell_3&2\cr 1&1&\lambda}\Bigr\rbrace
\cr
&
{{\sqrt{4\pi}}\over{\hat \lambda}}(-)^{k_{a_1}}
<j_{a_1}\Vert [Y^{(\ell_1)}\sigma]^{\lambda} \Vert j_{a_2}>
{{\sqrt{4\pi}}\over{\hat \lambda}}(-)^{k_{b_2}}
<j_{b_2}\Vert [Y^{(\ell_3)}\sigma]^{\lambda} \Vert j_{b_1}>
\cr
&~~\bar R_{a_1}(r_i)\bar R_{a_2}(r_i)\bar R_{b_1}(r_j)
   \bar R_{b_2}(r_j)
}\eqno{(2.41)}$$
The remaining term is:
$$\eqalign{
<(a_1\bar a_2)_{\lambda}&|V^{eff,k=1}|(b_2\bar b_1)_{\lambda}>=
72 A_{2\pi}~<\vec\tau_2\vec\tau_3>~~\cr &
 \hat \ell_1\hat \ell
\hat \ell_1\hat \ell 
<\ell_1 0\ell 0|20>^2
(-)^{(\ell+\ell_1+1)}
\Bigl\lbrace\matrix{\ell&\ell&1\cr 1&1&\lambda}\Bigr\rbrace
\Bigl\lbrace\matrix{\ell&\ell&1\cr 2&2&\ell_1}\Bigr\rbrace
\Bigl\lbrace\matrix{2&2&1\cr 1&1&1}\Bigr\rbrace\cr
&
{{\sqrt{4\pi}}\over{\hat \lambda}}(-)^{k_{a_1}}
<j_{a_1}\Vert [Y^{(\ell)}\sigma]^{\lambda} \Vert j_{a_2}>
{{\sqrt{4\pi}}\over{\hat \lambda}}(-)^{k_{b_2}}
<j_{b_2}\Vert [Y^{(\ell)}\sigma]^{\lambda} \Vert j_{b_1}>
\cr
&D^{\ell_1}_{m,s}V^{2,\ell_1\ell}_{m,n}
V^{2,\ell_1\ell}_{s,t}
\bar H^{\ell}_n(r_i)r_i^2w_i\bar H^{\ell}_t(r_j)r_j^2w_j
~~~\bar R_{a_1}(r_i)\bar R_{a_2}(r_i)\bar R_{b_1}(r_j)
   \bar R_{b_2}(r_j)
}\eqno{(2.41)}$$

\beginsection{3. Exchange matrix elements.}

In this section we work out the exchange matrix element which we define
as
$$V^{tni,x}_{p_1h_1;h_2p_2}=-
V^{tni,a}_{h,p_1,p_2;h_1,p,h_2}\eqno (3.1) $$
We assume the interaction can be written as sum over terms each having
the form:
$$-\Bigl[ T_1^{(\lambda_1)}(1)\otimes\bigl[ T_2^{(\lambda_2)}\otimes 
T_3^{(\lambda_3)} \bigr]^{(\lambda_1)}\Bigr]^{(0)}\eqno (3.2) $$
This leads to the exchange matrix element as
$$\eqalign{
V^{tni,x}_{p_1h_1;h_2p_2}=-\bigl\lbrace
&<h(1)p_1(2)p_2(3)|
\Bigl[ T_1^{(\lambda_1)}(1)\otimes\bigl[ T_2^{(\lambda_2)}\otimes 
T_3^{(\lambda_3)} \bigr]^{(\lambda_1)}\Bigr]^{(0)}
|h_1(1)p(2)h_2(3)>\cr
+&<h(2)p_1(3)p_2(1)|
\Bigl[ T_1^{(\lambda_1)}(1)\otimes\bigl[ T_2^{(\lambda_2)}\otimes 
T_3^{(\lambda_3)} \bigr]^{(\lambda_1)}\Bigr]^{(0)}
|h_1(2)p(3)h_2(1)>\cr
+&<h(3)p_1(1)p_2(2)|
\Bigl[ T_1^{(\lambda_1)}(1)\otimes\bigl[ T_2^{(\lambda_2)}\otimes 
T_3^{(\lambda_3)} \bigr]^{(\lambda_1)}\Bigr]^{(0)}
|h_1(3)p(1)h_2(2)> \bigr\rbrace }\eqno (3.3) $$
Here the three terms arise from the cyclic permutations.
Using our phase convention for $ph$-$ph$ angular momentum
coupling  given by Eqs. (1.4, 1.14) of Reference~[3], 
we can carry out the summation over all $m$'s except $m_p=m_h$,
using the Wigner-Eckart
theorem we obtain:
$$\eqalign{V^{tni,x,\lambda}_{p_1h_1,h_2p_2}=
(-)^{(k_{p_2}+k_{h_1}+\lambda_1+\lambda_2+\lambda)} 
{{\hat \lambda_1 \hat \lambda_2}\over{\hat \lambda}}
 \Bigl\lbrace\matrix{\lambda_1&\lambda_2&\lambda\cr j_{p_1}&
       j_{h_1}&j_h} \Bigr\rbrace 
\times \Bigl\lbrace 
& {1\over{\hat \lambda_1}}
\langle h\Vert T_1^{\lambda_1}\Vert h_1\rangle
 {1\over{\hat \lambda_2}}
\langle p_1\Vert T_2^{\lambda_2}\Vert h\rangle
 {1\over{\hat \lambda}}
\langle p_2\Vert T_3^{\lambda}\Vert h_2\rangle \cr
+ &{1\over{\hat \lambda}}
 \langle p_2\Vert T_2^{\lambda}\Vert h_2\rangle
 {1\over{\hat \lambda_1}}
\langle h\Vert T_3^{\lambda_1}\Vert h_1\rangle
 {1\over{\hat \lambda_2}}
\langle p_1\Vert T_1^{\lambda_2}\Vert h\rangle \cr
+ &{1\over{\hat \lambda_2}}
 \langle p_1\Vert T_3^{\lambda_2}\Vert h\rangle
 {1\over{\hat \lambda}}
\langle p_2\Vert T_1^{\lambda}\Vert h_2\rangle
 {1\over{\hat \lambda_1}}
\langle h\Vert T_2^{\lambda_1}\Vert h_1\rangle \Bigr\rbrace}\eqno (3.4)
$$

\beginsection{3a. Short range repulsion term.}

We first turn to the correction term due to the short range repulsion 
as the most simple contribution of $V^{tni}$. 
From section 2 we take the interaction as
$$\eqalign{
T^2(r_{12})T^2(r_{13})=&(4\pi)^2{1\over{\sqrt{4\pi}}}
\sum_{\ell_1,\ell_2,\ell3}
(-)^{(\ell_2+\ell_3)}\hat \ell_2 \hat \ell_3 
<\ell_2 0\ell_3 0|\ell_1 0>\cr
\times & {2\over{\pi}}\int q_2^2dq_2 \tilde T(q_2) 
j_{\ell_2}(q_2r_1) j_{\ell_2}(q_2r_2)
\times  {2\over{\pi}}\int q_3^2dq_3 \tilde T(q_3) 
j_{\ell_3}(q_3r_1) j_{\ell_3}(q_3r_3)\cr
\times & \Bigl[Y_{\ell_1}(\hat r_1)\otimes \bigl[Y_{\ell_2}(\hat
r_2)\otimes Y_{\ell_3}(\hat r_3)\bigr]^{(\ell_1)}\Bigr]^{(0)} 
}\eqno{(3.5)}$$
Combining this with Eq. (3.4) and considering the $\lambda$'s were
renamed in order to obtain (3.4) we find:
$$\eqalign{
V^{tni,x,\lambda}_{p_1h_1,h_2p_2}=& (-)^{(k_{p_1}+k_{h}+k_{h_1}+\lambda)}
{{(2\ell_1+1)(2\ell_2+1)}\over{\hat \lambda}}
\Bigl\lbrace\matrix{\ell_1&j_{h_1}&j_h\cr j_{p_1}&\ell_2&\lambda}
\Bigr\rbrace <\ell_1 0 \ell_2 0 | \lambda 0>\cr
\times  &
   \biggl[ (-)^{k_{p_2}}{\sqrt{4\pi}\over{\hat \lambda}}
   <p_2\Vert Y_{\lambda}\Vert h_2>\bigr]
   \bigl[ (-)^{k_{h}}{\sqrt{4\pi}\over{\hat \ell_1}}
   <h\Vert Y_{\ell_1}\Vert h_1>\bigr]
   \bigl[ (-)^{k_{p_1}}{\sqrt{4\pi}\over{\hat \ell_2}}
   <p_1\Vert Y_{\ell_2}\Vert p>\biggr]\cr
\times  \biggl\lbrace
&   \int r_1^2dr_1 R_h(r_1)R_{h_1}(r_1)\cr
& ~~~~~~~~~~ {2\over{\pi}}\int q_2^2dq_2 \tilde T(q_2)j_{\ell_2}(q_2r_1)
   \int r_2^2dr_2 R_{p_1}(r_2)R_p(r_2)j_{\ell_2}(q_2r_2)\cr
& ~~~~~~~~~~ {2\over{\pi}}\int q_3^2dq_3 \tilde T(q_3)j_{\lambda}(q_3r_1)
   \int r_3^2d3_2 R_{p_2}(r_3)R_{h_2}(r_3)j_{\lambda}(q_3r_3)\cr
}$$
\vfill
\eject
$$\eqalign{
+&   \int r_1^2dr_1 R_{p_2}(r_1)R_{h_2}(r_1)\cr
& ~~~~~~~~~~ {2\over{\pi}}\int q_2^2dq_2 \tilde T(q_2)j_{\ell_1}(q_2r_1)
   \int r_2^2dr_2 R_{h_1}(r_2)R_h(r_2)j_{\ell_1}(q_2r_2)\cr
& ~~~~~~~~~~ {2\over{\pi}}\int q_3^2dq_3 \tilde T(q_3)j_{\ell_2}(q_3r_1)
   \int r_3^2d3_2 R_{p_1}(r_3)R_{p}(r_3)j_{\ell_2}(q_3r_3)\cr
+&   \int r_1^2dr_1 R_{p_1}(r_1)R_{p}(r_1)\cr
& ~~~~~~~~~~ {2\over{\pi}}\int q_2^2dq_2 \tilde T(q_2)j_{\lambda}(q_2r_1)
   \int r_2^2dr_2 R_{p_2}(r_2)R_{h_2}(r_2)j_{\lambda}(q_2r_2)\cr
& ~~~~~~~~~~ {2\over{\pi}}\int q_3^2dq_3 \tilde T(q_3)j_{\ell_1}(q_3r_1)
   \int r_3^2d3_2 R_{h_1}(r_3)R_{h}(r_3)j_{\ell_1}(q_3r_3)
         \biggr\rbrace }\eqno (3.6) $$
Using Gauss integration, we write the radial matrix element as
$$\eqalign{
\Bigl\lbrace  r_k^2w_kR_h(r_k)R_{h_1}(r_k)& \bar H^{\ell_2}_m(r_k)
      T^{\ell_2}_{mn}\bar R_{p_1}(r_i)\bar R_p(r_i)\bar
      H^{\ell_2}_n(r_i) r_i^2w_i\cr
  & \bar H^{\lambda}_s(r_k) T^{\lambda}_{st} \bar R_{p_2}(r_j)
    \bar R_{h_2}(r_j) \bar H^{\lambda}_t(r_j) r_j^2 w_j\cr
  +  r_k^2w_kR_p(r_k)R_{p_1}(r_k)& \bar H^{\ell_1}_m(r_k)
      T^{\ell_1}_{mn}\bar R_{h_1}(r_i)\bar R_h(r_i)\bar
     H^{\ell_1}_n(r_i) r_i^2w_i\cr
  & \bar H^{\lambda}_s(r_k) T^{\lambda}_{st} \bar R_{p_2}(r_j)
    \bar R_{h_2}(r_j) \bar H^{\lambda}_t(r_j) r_j^2 w_j\cr
  +  r_k^2w_kR_{p_2}(r_k)R_{h_2}(r_k)& \bar H^{\ell_1}_m(r_k)
     T^{\ell_1}_{mn}\bar R_{h_1}(r_i)\bar R_h(r_i)\bar
     H^{\ell_1}_n(r_i) r_i^2w_i\cr
  & \bar H^{\ell_2}_s(r_k)T^{\ell_2}_{st}\bar R_{p_1}(r_j)\bar
     R_p(r_j)\bar H^{\ell_2}_t(r_j) r_j^2w_j \Bigr\rbrace }\eqno (3.7)$$
defining the kernel
$$G^{\ell}(r_k,r_i)=\sqrt{r_k^2w_k}\bar H^{\ell}_m(r_k)T^{\ell}_{mn}
   \bar H^{\ell}_n(r_i)r_i^2w_i\eqno (3.8)$$
allows us to write the radial integrals as
$$\eqalign{
\Bigl\lbrace  R_h(r_k)R_{h_1}(r_k)& 
      \bar R_{p_1}(r_i)\bar R_p(r_i)
      G^{\ell_2}(r_k,r_i)G^{\lambda}(r_k,r_j)
      \bar R_{p_2}(r_j) \bar R_{h_2}(r_j) \cr
  +  R_p(r_k)R_{p_1}(r_k)& 
      \bar R_{h_1}(r_i)\bar R_h(r_i)
     G^{\ell_1}(r_k,r_i)G^{\lambda}(r_k,r_j)
     \bar R_{p_2}(r_j) \bar R_{h_2}(r_j) \cr
  +  R_{p_2}(r_k)R_{h_2}(r_k)& 
     \bar R_{h_1}(r_i)\bar R_h(r_i)
     G^{\ell_1}(r_k,r_i)G^{\ell_2}(r_k,r_j)
    \bar R_{p_1}(r_j)\bar R_p(r_j)
     \Bigr\rbrace }\eqno (3.9)$$

\beginsection{3b. Two-pion exchange term.}

Again, we take the interaction from section 2 as
$$\eqalign{A_{2\pi}(4\pi)^{3/2}
\sum_{K_2=0,2}\sum_{K_3=0,2}&<1010|K_20><1010|K_30>
(-)^{\ell_2+\lambda_2+\ell_3+\lambda_3} \cr
&{2\over{\pi}}\int q_2^2dq_2 \tilde V^{K_2}(q_2)\hat \ell_1\hat\ell_2
(i)^{(\ell_1-\ell_2-K_2)}j_{\ell_1}(q_2r_1)j_{\ell_2}(q_2r_2)
{{3}\over{\hat K_2}}<\ell_1 0\ell_2 0|K_20>\cr
&{2\over{\pi}}\int q_3^2dq_3 \tilde V^{K_3}(q_3)\hat \ell_4\hat\ell_3
(i)^{(\ell_4-\ell_3-K_3)}j_{\ell_4}(q_3r_1)j_{\ell_3}(q_3r_3)
{{3}\over{\hat K_3}}<\ell_4 0\ell_3 0|K_30>\cr
&\hat K_2 \hat K_3 \hat
\lambda_2 \hat \lambda_3 \hat \lambda_1 \hat \ell_1 \hat \ell_4 \hat S
\Biggl\lbrace
\matrix{\ell_1 & 1 & \lambda_2\cr \ell_4& 1& \lambda_3\cr
L & S & \lambda_1}\Biggr\rbrace
\Bigl\lbrace\matrix{\ell_1&\ell_2&K_2\cr 1&1&\lambda_2}\Bigr\rbrace
\Bigl\lbrace\matrix{\ell_4&\ell_3&K_3\cr 1&1&\lambda_3}\Bigr\rbrace
\langle \ell_1 0\ell_4 0|L0\rangle\cr
& \Biggl[\biggl[Y_{L}(\hat r_1)
     \otimes  \bigl[\sigma_1\otimes\sigma_1\bigr]
   ^{(S)}\biggr]^{(\lambda_1)} \otimes
     \biggl[\bigl[Y_{\ell_2}(\hat
     r_2)\otimes\sigma_2\bigr]^{(\lambda_2)}\otimes \bigl[Y_{\ell_3}(\hat
     r_3)\otimes\sigma_3\bigr]^{(\lambda_3)}\biggr]^{(\lambda_1)}\Biggr]^{(0)}
}\eqno{(3.8)}$$
Here $K_2$ or $K_3$ are either 0, for the $\sigma\cdot\sigma$-term
or 2, for the tensor term.  Further, $S$ is 0 in the anti-commutator
term and 1 in the commutator term. As for the present correction we only
have a contribution from the anticommutator, we write the interaction
as twice the $S$=0 contribution:
$$\eqalign{
18A_{2\pi}&(4\pi)^{3/2}
\sum_{K_2=0,2}\sum_{K_3=0,2}<1010|K_20><1010|K_30>
(-)^{\lambda_2+\ell_3+\lambda_1}\cr
&{2\over{\pi}}\int q_2^2dq_2 \tilde V^{K_2}(q_2)\hat \ell_1\hat\ell_2
(i)^{(\ell_1-\ell_2-K_2)}j_{\ell_1}(q_2r_1)j_{\ell_2}(q_2r_2)
<\ell_1 0\ell_2 0|K_20>\cr
&{2\over{\pi}}\int q_3^2dq_3 \tilde V^{K_3}(q_3)\hat \ell_4\hat\ell_3
(i)^{(\ell_4-\ell_3-K_3)}j_{\ell_4}(q_3r_1)j_{\ell_3}(q_3r_3)
<\ell_4 0\ell_3 0|K_30>\cr
& \hat \lambda_2 \hat \lambda_3 \hat \ell_1 \hat \ell_4
\Bigl\lbrace\matrix{\lambda_2&\ell_1&1\cr 
\ell_4&\lambda_3&\lambda_1}\Bigr\rbrace
\Bigl\lbrace\matrix{\ell_1&\ell_2&K_2\cr 1&1&\lambda_2}\Bigr\rbrace
\Bigl\lbrace\matrix{\ell_4&\ell_3&K_3\cr 1&1&\lambda_3}\Bigr\rbrace
\langle \ell_1 0\ell_4 0|\lambda_10\rangle\cr
&\Biggl[Y_{\lambda_1}(\hat r_1)
    \otimes
     \biggl[\bigl[Y_{\ell_2}(\hat
     r_2)\otimes\sigma_2\bigr]^{(\lambda_2)}\otimes \bigl[Y_{\ell_3}(\hat
     r_3)\otimes\sigma_3\bigr]^{(\lambda_3)}\biggr]^{(\lambda_1)}\Biggr]^{(0)}
}\eqno{(3.9)}$$
We carry out the summation over all $m$'s and do the summation over
cyclic permutations by using Eq. (3.4). However, it should be noted that
in the second and third term of the cyclic permutations the
$\lambda$'s in Eq. (3.4) were renamed
$$\eqalign{
18A_{2\pi}&<1010|K_20><1010|K_30>\hat \ell_1 \hat \ell_4
\hat \ell_1\hat\ell_2\hat \ell_4\hat\ell_3{{\hat \lambda_1 \hat
\lambda_2}\over{\hat \lambda}}
 \Bigl\lbrace\matrix{\lambda_1&\lambda_2&\lambda\cr j_{p_1}&
       j_{h_1}&j_h} \Bigr\rbrace
(-)^{\ell_3}\cr
& (i)^{(\ell_1-\ell_2-K_2+
\ell_4-\ell_3-K_3)}
<\ell_1 0\ell_2 0|K_20>
<\ell_4 0\ell_3 0|K_30>(-)^{(k_h+k_{p_1}+k_{h_1})}\cr
&{2\over{\pi}}\int q_2^2dq_2 \tilde V^{K_2}(q_2)
j_{\ell_1}(q_2r_1)j_{\ell_2}(q_2r_2)
\cr
&{2\over{\pi}}\int q_3^2dq_3 \tilde V^{K_3}(q_3)
j_{\ell_4}(q_3r_1)j_{\ell_3}(q_3r_3)
\cr
\biggl\lbrace
& \hat \lambda_2 \hat \lambda
\Bigl\lbrace\matrix{\lambda_2&\ell_1&1\cr
\ell_4&\lambda&\lambda_1}\Bigr\rbrace
\Bigl\lbrace\matrix{\ell_1&\ell_2&K_2\cr 1&1&\lambda_2}\Bigr\rbrace
\Bigl\lbrace\matrix{\ell_4&\ell_3&K_3\cr 1&1&\lambda}\Bigr\rbrace
\langle \ell_1 0\ell_4 0|\lambda_10\rangle (-)^{\lambda}
\cr
&\bigl[(-)^{k_h}{{\sqrt{4\pi}}\over{\hat \lambda_1}}
 \langle h\Vert Y^{\lambda_1}\Vert h_1\rangle\bigr]
 \bigl[(-)^{k_{p_1}}{{\sqrt{4\pi}}\over{\hat \lambda_2}}
\langle p_1\Vert [Y_{\ell_2}\sigma]^{\lambda_2}\Vert p\rangle\bigr]
 \bigl[(-)^{k_{p_2}}{{\sqrt{4\pi}}\over{\hat \lambda}}
 \langle p_2\Vert [Y_{\ell_3}\sigma]^{\lambda}\Vert h_2\rangle \bigr]\cr
+
& \hat \lambda_1 \hat \lambda_2
\Bigl\lbrace\matrix{\lambda_1&\ell_1&1\cr
\ell_4&\lambda_2&\lambda}\Bigr\rbrace
\Bigl\lbrace\matrix{\ell_1&\ell_2&K_2\cr 1&1&\lambda_1}\Bigr\rbrace
\Bigl\lbrace\matrix{\ell_4&\ell_3&K_3\cr 1&1&\lambda_2}\Bigr\rbrace
\langle \ell_1 0\ell_4 0|\lambda 0\rangle (-)^{\lambda_2}
\cr
&\bigl[(-)^{k_{p_2}}{{\sqrt{4\pi}}\over{\hat \lambda}}
 \langle p_2\Vert Y^{\lambda}\Vert h_2\rangle\bigr]
 \bigl[(-)^{k_h}{{\sqrt{4\pi}}\over{\hat \lambda_1}}
 \langle h\Vert [Y_{\ell_2}\sigma]^{\lambda_1}\Vert h_1\rangle\bigr]
 \bigl[(-)^{k_{p_1}}{{\sqrt{4\pi}}\over{\hat \lambda_2}}
 \langle p_1\Vert [Y_{\ell_3}\sigma]^{\lambda_2}\Vert p\rangle
\bigr]\cr
}$$
\vfill
\eject
$$\eqalign{
+
& \hat \lambda \hat \lambda_1
\Bigl\lbrace\matrix{\lambda&\ell_1&1\cr
\ell_4&\lambda_1&\lambda_2}\Bigr\rbrace
\Bigl\lbrace\matrix{\ell_1&\ell_2&K_2\cr 1&1&\lambda}\Bigr\rbrace
\Bigl\lbrace\matrix{\ell_4&\ell_3&K_3\cr 1&1&\lambda_1}\Bigr\rbrace
\langle \ell_1 0\ell_4 0|\lambda_20\rangle (-)^{\lambda_1}
\cr
&\bigl[(-)^{k_{p_1}}{{\sqrt{4\pi}}\over{\hat \lambda_2}}
 \langle p_1\Vert Y^{\lambda_2}\Vert p\rangle\bigr]
 \bigl[(-)^{k_{p_2}}{{\sqrt{4\pi}}\over{\hat \lambda}}
 \langle p_2\Vert [Y_{\ell_2}\sigma]^{\lambda}\Vert h_2\rangle\bigr]
 \bigl[(-)^{k_h}{{\sqrt{4\pi}}\over{\hat \lambda_1}}
 \langle h\Vert [Y_{\ell_3}\sigma]^{\lambda_1}\Vert h_1\rangle
\bigr]\biggr\rbrace }\eqno(3.10)$$
We write the radial integrals as summations over the Gauss points.
We compute the matrix element as
$$\eqalign{
V^{tni,x,\lambda}_{p_1h_1,h_2p_2}=& 
18A_{2\pi} <1010|K_20><1010|K_30>\hat \ell_1 \hat \ell_4
\hat \ell_1\hat\ell_2\hat \ell_4\hat\ell_3{{\hat \lambda_1 \hat
\lambda_2}\over{\hat \lambda}}
 \Bigl\lbrace\matrix{\lambda_1&\lambda_2&\lambda\cr j_{p_1}&
       j_{h_1}&j_h} \Bigr\rbrace 
\cr
& (i)^{(\ell_1-\ell_2-K_2+
\ell_4-\ell_3-K_3)}
<\ell_1 0\ell_2 0|K_20>
<\ell_4 0\ell_3 0|K_30>(-)^{(k_h+k_{p_1}+k_{h_1})}\cr
\biggl\lbrace 
& \hat \lambda_2 \hat \lambda 
\Bigl\lbrace\matrix{\lambda_2&\ell_1&1\cr 
\ell_4&\lambda&\lambda_1}\Bigr\rbrace
\Bigl\lbrace\matrix{\ell_1&\ell_2&K_2\cr 1&1&\lambda_2}\Bigr\rbrace
\Bigl\lbrace\matrix{\ell_4&\ell_3&K_3\cr 1&1&\lambda}\Bigr\rbrace
\langle \ell_1 0\ell_4 0|\lambda_10\rangle (-)^{\lambda+\ell_3}
\cr
&\bigl[(-)^{k_h}{{\sqrt{4\pi}}\over{\hat \lambda_1}}
 \langle h\Vert Y^{\lambda_1}\Vert h_1\rangle\bigr]
 \bigl[(-)^{k_{p_1}}{{\sqrt{4\pi}}\over{\hat \lambda_2}}
\langle p_1\Vert [Y_{\ell_2}\sigma]^{\lambda_2}\Vert p\rangle\bigr]
 \bigl[(-)^{k_{p_2}}{{\sqrt{4\pi}}\over{\hat \lambda}}
 \langle p_2\Vert [Y_{\ell_3}\sigma]^{\lambda}\Vert h_2\rangle \bigr]\cr
 &r_k^2w_kR_h(r_k)R_{h_1}(r_k) \bar H^{\ell_1}_m(r_k)
      T^{K_2,\ell_1,\ell_2}_{mn}\bar R_{p_1}(r_i)\bar R_p(r_i)\bar
      H^{\ell_2}_n(r_i) r_i^2w_i\cr
  &~~~~~ \bar H^{\ell_4}_s(r_k) T^{K_3,\ell_4,\ell_3}_{st} \bar R_{p_2}(r_j)
    \bar R_{h_2}(r_j) \bar H^{\ell_3}_t(r_j) r_j^2 w_j\cr
+ 
& \hat \lambda_1 \hat \lambda_2 
\Bigl\lbrace\matrix{\lambda_1&\ell_1&1\cr 
\ell_4&\lambda_2&\lambda}\Bigr\rbrace
\Bigl\lbrace\matrix{\ell_1&\ell_2&K_2\cr 1&1&\lambda_1}\Bigr\rbrace
\Bigl\lbrace\matrix{\ell_4&\ell_3&K_3\cr 1&1&\lambda_2}\Bigr\rbrace
\langle \ell_1 0\ell_4 0|\lambda 0\rangle (-)^{\lambda_2+\ell_3}
\cr
&\bigl[(-)^{k_{p_2}}{{\sqrt{4\pi}}\over{\hat \lambda}}
 \langle p_2\Vert Y^{\lambda}\Vert h_2\rangle\bigr]
 \bigl[(-)^{k_h}{{\sqrt{4\pi}}\over{\hat \lambda_1}}
 \langle h\Vert [Y_{\ell_2}\sigma]^{\lambda_1}\Vert h_1\rangle\bigr]
 \bigl[(-)^{k_{p_1}}{{\sqrt{4\pi}}\over{\hat \lambda_2}}
 \langle p_1\Vert [Y_{\ell_3}\sigma]^{\lambda_2}\Vert p\rangle 
\bigr]\cr
  & r_k^2w_kR_{p_2}(r_k)R_{h_2}(r_k) \bar H^{\ell_1}_m(r_k)
      T^{K_2,\ell_1,\ell_2}_{mn}\bar R_{h_1}(r_i)\bar R_h(r_i)\bar
     H^{\ell_2}_n(r_i) r_i^2w_i\cr
  &~~~~~ \bar H^{\ell_4}_s(r_k) T^{K_3,\ell_4\ell_3}_{st} \bar R_{p_1}(r_j)
    \bar R_{p}(r_j) \bar H^{\ell_3}_t(r_j) r_j^2 w_j\cr
+ 
& \hat \lambda \hat \lambda_1 
\Bigl\lbrace\matrix{\lambda&\ell_1&1\cr 
\ell_4&\lambda_1&\lambda_2}\Bigr\rbrace
\Bigl\lbrace\matrix{\ell_4&\ell_3&K_3\cr 1&1&\lambda_1}\Bigr\rbrace
\Bigl\lbrace\matrix{\ell_1&\ell_2&K_2\cr 1&1&\lambda}\Bigr\rbrace
\langle \ell_1 0\ell_4 0|\lambda_20\rangle (-)^{\lambda_1+\ell_3}
\cr
&\bigl[(-)^{k_{p_1}}{{\sqrt{4\pi}}\over{\hat \lambda_2}}
 \langle p_1\Vert Y^{\lambda_2}\Vert p\rangle\bigr]
 \bigl[(-)^{k_{p_2}}{{\sqrt{4\pi}}\over{\hat \lambda}}
 \langle p_2\Vert [Y_{\ell_2}\sigma]^{\lambda}\Vert h_2\rangle\bigr]
 \bigl[(-)^{k_h}{{\sqrt{4\pi}}\over{\hat \lambda_1}}
 \langle h\Vert [Y_{\ell_3}\sigma]^{\lambda_1}\Vert h_1\rangle\bigr] \cr
  &  r_k^2w_kR_{p_1}(r_k)R_{p}(r_k) \bar H^{\ell_4}_m(r_k)
     T^{K_3,\ell_4,\ell_3}_{mn}\bar R_{h_1}(r_i)\bar R_h(r_i)\bar
     H^{\ell_3}_n(r_i) r_i^2w_i\cr
  & ~~~~~\bar H^{\ell_1}_s(r_k)T^{K_2,\ell_1,\ell_2}_{st}\bar R_{p_2}(r_j)\bar
     R_{h_2}(r_j)\bar H^{\ell_2}_t(r_j) r_j^2w_j 
\biggr\rbrace }\eqno(3.11)$$
We define the interaction kernel as:
$$\eqalign{
K^{\lambda,\ell_1\ell_2}(r_k,r_i):= &\sum_K
 (i)^{(\ell_1-\ell_2-K)}
 <1010|K0>\hat \ell_1 
\hat \ell_1\hat\ell_2
<\ell_1 0\ell_2 0|K0>
\Bigl\lbrace\matrix{\ell_1&\ell_2&K\cr 1&1&\lambda}\Bigr\rbrace
\hat \lambda \cr
 &\sum_{m,n} \sqrt{r_k^2w_k} \bar H^{\ell_1}_m(r_k)
      T^{K,\ell_1,\ell_2}_{mn} H^{\ell_2}_n(r_i) r_i^2w_i}\eqno(3.12)$$
With this definition we write the matrix element as
$$\eqalign{
V^{tni,x,\lambda}_{p_1h_1,h_2p_2}=&
18A_{2\pi} {{\hat \lambda_1 \hat
\lambda_2}\over{\hat \lambda}}(-)^{(k_h+k_{p_1}+k_{h_1})}
 \Bigl\lbrace\matrix{\lambda_1&\lambda_2&\lambda\cr j_{p_1}&
       j_{h_1}&j_h} \Bigr\rbrace \cr
\biggl\lbrace
& \Bigl\lbrace\matrix{\lambda_2&\ell_1&1\cr
\ell_4&\lambda&\lambda_1}\Bigr\rbrace
\langle \ell_1 0\ell_4 0|\lambda_10\rangle (-)^{\lambda+\ell_3}
\cr
&\bigl[(-)^{k_h}{{\sqrt{4\pi}}\over{\hat \lambda_1}}
 \langle h\Vert Y^{\lambda_1}\Vert h_1\rangle\bigr]
 \bigl[(-)^{k_{p_1}}{{\sqrt{4\pi}}\over{\hat \lambda_2}}
\langle p_1\Vert [Y_{\ell_2}\sigma]^{\lambda_2}\Vert p\rangle\bigr]
 \bigl[(-)^{k_{p_2}}{{\sqrt{4\pi}}\over{\hat \lambda}}
 \langle p_2\Vert [Y_{\ell_3}\sigma]^{\lambda}\Vert h_2\rangle \bigr]\cr
&R_h(r_k)R_{h_1}(r_k)K^{\lambda_2,\ell_1,\ell_2}(r_k,r_i)\bar
R_{p_1}(r_i)\bar R_p(r_i)K^{\lambda,\ell_4,\ell_3}(r_k,r_j)
\bar R_{p_2}(r_j) \bar R_{h_2}(r_j)\cr
}$$
\vfill
\eject
$$\eqalign{
+
& \Bigl\lbrace\matrix{\lambda_1&\ell_1&1\cr
\ell_4&\lambda_2&\lambda}\Bigr\rbrace
\langle \ell_1 0\ell_4 0|\lambda 0\rangle (-)^{\lambda_2+\ell_3}\cr
&\bigl[(-)^{k_{p_2}}{{\sqrt{4\pi}}\over{\hat \lambda}}
 \langle p_2\Vert Y^{\lambda}\Vert h_2\rangle\bigr]
 \bigl[(-)^{k_h}{{\sqrt{4\pi}}\over{\hat \lambda_1}}
 \langle h\Vert [Y_{\ell_2}\sigma]^{\lambda_1}\Vert h_1\rangle\bigr]
 \bigl[(-)^{k_{p_1}}{{\sqrt{4\pi}}\over{\hat \lambda_2}}
 \langle p_1\Vert [Y_{\ell_3}\sigma]^{\lambda_2}\Vert p\rangle
\bigr]\cr
&R_{p_2}(r_k)R_{h_2}(r_k)K^{\lambda_1,\ell_1,\ell_2}(r_k,r_i)
\bar R_{h_1}(r_i)\bar R_h(r_i)K^{\lambda_2,\ell_4,\ell_3}(r_k,r_j)
 \bar R_{p_1}(r_j) \bar R_{p}(r_j) \cr
+
& \Bigl\lbrace\matrix{\lambda&\ell_1&1\cr
\ell_4&\lambda_1&\lambda_2}\Bigr\rbrace
\langle \ell_1 0\ell_4 0|\lambda_20\rangle (-)^{\lambda_1+\ell_3}\cr
&\bigl[(-)^{k_{p_1}}{{\sqrt{4\pi}}\over{\hat \lambda_2}}
 \langle p_1\Vert Y^{\lambda_2}\Vert p\rangle\bigr]
 \bigl[(-)^{k_{p_2}}{{\sqrt{4\pi}}\over{\hat \lambda}}
 \langle p_2\Vert [Y_{\ell_2}\sigma]^{\lambda}\Vert h_2\rangle\bigr]
 \bigl[(-)^{k_h}{{\sqrt{4\pi}}\over{\hat \lambda_1}}
 \langle h\Vert [Y_{\ell_3}\sigma]^{\lambda_1}\Vert h_1\rangle\bigr] \cr
& R_{p_1}(r_k)R_{p}(r_k)K^{\lambda_1,\ell_4\ell_3}(r_k,r_i)
 \bar R_{h_1}(r_i)\bar R_h(r_i)K^{\lambda,\ell_1,\ell_2}(r_k,r_j)
\bar R_{p_2}(r_j)\bar R_{h_2}(r_j)\biggr\rbrace }\eqno(3.13)$$
Similarly, we get the commutator contribution as
$$\eqalign{
V^{tni,cx,\lambda}_{p_1h_1,h_2p_2}=&
{9 \over 2} \sqrt{{3\over2}} A_{2\pi} {{\hat \lambda_1 \hat
\lambda_2}\over{\hat \lambda}}(-)^{(k_h+k_{p_1}+k_{h_1})}
 \Bigl\lbrace\matrix{\lambda_1&\lambda_2&\lambda\cr j_{p_1}&
       j_{h_1}&j_h} \Bigr\rbrace \cr
\Biggl\lbrace
& \Biggl\lbrace\matrix{\ell_1&1&\lambda_2\cr
                       \ell_4&1&\lambda\cr
                       L&1&\lambda_1
                      }\Biggr\rbrace \hat \lambda_1
\langle \ell_1 0\ell_4 0|L 0\rangle (-)^{\lambda_1+\ell_2+\ell_3}
\cr
&\bigl[(-)^{k_h}{{\sqrt{4\pi}}\over{\hat \lambda_1}}
 \langle h\Vert [Y_{L}\sigma]^{\lambda_1}\Vert h_1\rangle\bigr]
 \bigl[(-)^{k_{p_1}}{{\sqrt{4\pi}}\over{\hat \lambda_2}}
\langle p_1\Vert [Y_{\ell_2}\sigma]^{\lambda_2}\Vert p\rangle\bigr]
 \bigl[(-)^{k_{p_2}}{{\sqrt{4\pi}}\over{\hat \lambda}}
 \langle p_2\Vert [Y_{\ell_3}\sigma]^{\lambda}\Vert h_2\rangle \bigr]\cr
&R_h(r_k)R_{h_1}(r_k)K^{\lambda_2,\ell_1,\ell_2}(r_k,r_i)\bar
R_{p_1}(r_i)\bar R_p(r_i)K^{\lambda,\ell_4,\ell_3}(r_k,r_j)
\bar R_{p_2}(r_j) \bar R_{h_2}(r_j)\cr
+
& \Biggl\lbrace\matrix{\ell_1&1&\lambda_1\cr
                       \ell_4&1&\lambda_2\cr
                       L&1&\lambda
                      }\Biggr\rbrace \hat \lambda
\langle \ell_1 0\ell_4 0|L 0\rangle (-)^{\lambda+\ell_2+\ell_3}\cr
&\bigl[(-)^{k_{p_2}}{{\sqrt{4\pi}}\over{\hat \lambda}}
 \langle p_2\Vert [Y_{L}\sigma]^{\lambda}\Vert h_2\rangle\bigr]
 \bigl[(-)^{k_h}{{\sqrt{4\pi}}\over{\hat \lambda_1}}
 \langle h\Vert [Y_{\ell_2}\sigma]^{\lambda_1}\Vert h_1\rangle\bigr]
 \bigl[(-)^{k_{p_1}}{{\sqrt{4\pi}}\over{\hat \lambda_2}}
 \langle p_1\Vert [Y_{\ell_3}\sigma]^{\lambda_2}\Vert p\rangle
\bigr]\cr
&R_{p_2}(r_k)R_{h_2}(r_k)K^{\lambda_1,\ell_1,\ell_2}(r_k,r_i)
\bar R_{h_1}(r_i)\bar R_h(r_i)K^{\lambda_2,\ell_4,\ell_3}(r_k,r_j)
 \bar R_{p_1}(r_j) \bar R_{p}(r_j) \cr
+
& \Biggl\lbrace\matrix{\ell_1&1&\lambda\cr
                       \ell_4&1&\lambda_1\cr
                       L&1&\lambda_2
                       }\Biggr\rbrace \hat \lambda_2
\langle \ell_1 0\ell_4 0|L 0\rangle (-)^{\lambda_2+\ell_2+\ell_3}\cr
&\bigl[(-)^{k_{p_1}}{{\sqrt{4\pi}}\over{\hat \lambda_2}}
 \langle p_1\Vert [Y_{L}\sigma]^{\lambda_2}\Vert p\rangle\bigr]
 \bigl[(-)^{k_{p_2}}{{\sqrt{4\pi}}\over{\hat \lambda}}
 \langle p_2\Vert [Y_{\ell_2}\sigma]^{\lambda}\Vert h_2\rangle\bigr]
 \bigl[(-)^{k_h}{{\sqrt{4\pi}}\over{\hat \lambda_1}}
 \langle h\Vert [Y_{\ell_3}\sigma]^{\lambda_1}\Vert h_1\rangle\bigr] \cr
& R_{p_1}(r_k)R_{p}(r_k)K^{\lambda_1,\ell_4\ell_3}(r_k,r_i)
 \bar R_{h_1}(r_i)\bar R_h(r_i)K^{\lambda,\ell_1,\ell_2}(r_k,r_j)
\bar R_{p_2}(r_j)\bar R_{h_2}(r_j)\Biggr\rbrace }\eqno(3.14)$$

\beginsection{References}

\frenchspacing 

\item{[1]}{ J.~H.~Heisenberg and B.~Mihaila, ``Ground state
correlations and mean-field in $^{16}$O: Part II,'' nucl-th/9912023
(1998).}

\item{[2]}{ J.~Carlson, V.R.~Pandharipande, and R.B.~Wiringa,
Nucl. Phys. A \bf{401}, 59 (1983).}

\item{[3]}{ B.~Mihaila and J.~H.~Heisenberg, 'Computation of two-body
matrix elements from the Argonne V-18 potential', nucl-th/9802012
(1998).}

\vfill
\eject
\bye